\documentclass[twocolumn]{aastex631}

\newcommand{\teff}{\ensuremath{T_{\rm eff}}}

\newcommand{\msun}{\ensuremath{\,M_\Sun}}
\newcommand{\rsun}{\ensuremath{\,R_\Sun}}

\newcommand{\mstar}{\ensuremath{M_{*}}}

\newcommand{\degree}{\ensuremath{\,^{\circ}}}
\newcommand{\prot}{$P_{\rm rot}$}

\usepackage{url}
\usepackage{graphicx}
\usepackage{amsmath,amstext}
\usepackage[T1]{fontenc}

\usepackage{amsmath}
\usepackage{float}
\usepackage{longtable}
\usepackage{rotating}
\usepackage{multirow}

\graphicspath{{./}{Figures/}}

\shorttitle{Near-Resonant \& Near-Aligned}
\shortauthors{Rice et al.}

\begin{document}
\title{Evidence for Low-Level Dynamical Excitation in Near-Resonant Exoplanet Systems\footnote{This paper includes data gathered with the 6.5 meter Magellan Telescopes located at Las Campanas Observatory, Chile.}}

% Author order not final 

\author[0000-0002-7670-670X]{Malena Rice} %confirmed
\affiliation{Department of Astronomy, Yale University, New Haven, CT 06511, USA}
\affiliation{Department of Physics and Kavli Institute for Astrophysics and Space Research, Massachusetts Institute of Technology, Cambridge, MA 02139, USA}

\author[0000-0002-0376-6365]{Xian-Yu Wang} %confirmed
\affiliation{Department of Astronomy, Indiana University, Bloomington, IN 47405, USA}

\author[0000-0002-7846-6981]{Songhu Wang} %confirmed
\affiliation{Department of Astronomy, Indiana University, Bloomington, IN 47405, USA}

% co-author on proposal and helped to organize observations
\author[0000-0002-1836-3120]{Avi Shporer} %confirmed
\affiliation{Department of Physics and Kavli Institute for Astrophysics and Space Research, Massachusetts Institute of Technology, Cambridge, MA 02139, USA}

% contributed observations
\author[0000-0003-1464-9276]{Khalid Barkaoui} %confirmed
\affiliation{Astrobiology Research Unit, University of Li\`ege, All\'ee du 6 ao\^ut, 19, 4000 Li\`ege (Sart-Tilman), Belgium}
\affiliation{Department of Earth, Atmospheric and Planetary Sciences, MIT, 77 Massachusetts Avenue, Cambridge, MA 02139, USA}
\affiliation{Instituto de Astrof\'isica de Canarias (IAC), Calle V\'ia L\'actea s/n, 38200, La Laguna, Tenerife, Spain}

\author[0000-0002-9158-7315]{Rafael Brahm} %confirmed
\affiliation{Millennium Institute for Astrophysics, Chile}
\affiliation{Facultad de Ingeniera y Ciencias, Universidad Adolfo Ibáñez, Av. Diagonal las Torres 2640, Peñalolén, Santiago, Chile}
\affiliation{Data Observatory Foundation, Chile}

\author[0000-0001-6588-9574]{Karen A. Collins} %confirmed
\affiliation{Center for Astrophysics, Harvard $\&$ Smithsonian, 60 Garden Street, Cambridge, MA 02138, USA}

\author[0000-0002-5389-3944]{Andrés Jordán} %confirmed
\affiliation{Millennium Institute for Astrophysics, Chile}
\affiliation{Facultad de Ingeniera y Ciencias, Universidad Adolfo Ibáñez, Av. Diagonal las Torres 2640, Peñalolén, Santiago, Chile}
\affiliation{Data Observatory Foundation, Chile}

\author[0000-0001-6508-5736]{Nataliea Lowson} %confirmed
\affiliation{Centre for Astrophysics, University of Southern Queensland, 499-565 West Street, Toowoomba, QLD 4350, Australia}

%\author[0000-0002-4891-3517]{George Zhou}
%\affiliation{University of Southern Queensland, Centre for Astrophysics, West Street, Toowoomba, QLD 4350, Australia}

% PFS team 
\author[0000-0003-1305-3761]{R. Paul Butler} %confirmed
\affiliation{Carnegie Institution for Science, Earth \& Planets Laboratory, 5241 Broad Branch Road NW, Washington, DC 20015, USA}

\author[0000-0002-5226-787X]{Jeffrey D. Crane} %confirmed
\affiliation{The Observatories of the Carnegie Institution for Science, 813 Santa Barbara Street, Pasadena, CA 91101, USA}

\author[0000-0002-8681-6136]{Stephen Shectman} %confirmed
\affiliation{The Observatories of the Carnegie Institution for Science, 813 Santa Barbara Street, Pasadena, CA 91101, USA}

\author[0009-0008-2801-5040]{Johanna K. Teske} %confirmed
\affiliation{Carnegie Institution for Science, Earth \& Planets Laboratory, 5241 Broad Branch Road NW, Washington, DC 20015, USA}

%\author{Ian Thompson}
%\affiliation{The Observatories of the Carnegie Institution for Science, 813 Santa Barbara Street, Pasadena, CA 91101, USA}

\author[0000-0003-0412-9664]{David Osip}
\affiliation{Las Campanas Observatory, Carnegie Institution for Science, COlina el Pino, Casilla 601 La Serena, Chile}

% LCO coauthors 

%kcolli3@gmu.edu  (LCO observation coordination)   
\author[0000-0003-2781-3207]{Kevin I.\ Collins} %confirmed
\affiliation{George Mason University, 4400 University Drive, Fairfax, VA, 22030 USA}

%fmurgas@iac.es    (LCO 0m4 time contribution from IAC)
\author[0000-0001-9087-1245]{Felipe Murgas}
\affiliation{Instituto de Astrof\'isica de Canarias (IAC), E-38205 La Laguna, Tenerife, Spain}
\affiliation{Departamento de Astrof\'isica, Universidad de La Laguna (ULL), E-38206 La Laguna, Tenerife, Spain}

% Observatoire Moana 
\author[0009-0009-2966-7507]{Gavin Boyle} % gavinsboyle@icloud.com; confirmed
\affiliation{El Sauce Observatory -- Obstech, Chile}
\affiliation{Cavendish Laboratory, JJ Thomson Avenue, Cambridge, CB3 0HE, UK}

% TRAPPIST-South team 

\author[0000-0003-1572-7707]{Francisco J. Pozuelos} %confirmed
\affiliation{Instituto de Astrof\'isica de Andaluc\'ia (IAA-CSIC), Glorieta de la Astronom\'ia s/n, 18008 Granada, Spain}
\affiliation{Astrobiology Research Unit, University of Li\`ege, All\'ee du 6 ao\^ut, 19, 4000 Li\`ege (Sart-Tilman), Belgium}

\author{Mathilde Timmermans} %confirmed
\affiliation{Astrobiology Research Unit, University of Li\`ege, All\'ee du 6 ao\^ut, 19, 4000 Li\`ege (Sart-Tilman), Belgium}

\author[0000-0001-8923-488X]{Emmanuel Jehin} %confirmed
\affiliation{Space Sciences, Technologies and Astrophysics Research (STAR) Institute, Universit\'e de Li\`ege, All\'ee du 6 Ao\^ut 19C, B-4000 Li\`ege, Belgium}

\author[0000-0003-1462-7739]{Micha\"el Gillon} %confirmed
\affiliation{Astrobiology Research Unit, University of Li\`ege, All\'ee du 6 ao\^ut, 19, 4000 Li\`ege (Sart-Tilman), Belgium}

\correspondingauthor{Malena Rice}
\email{malena.rice@yale.edu}

\begin{abstract}
The geometries of near-resonant planetary systems offer a relatively pristine window into the initial conditions of exoplanet systems. Given that near-resonant systems have likely experienced minimal dynamical disruptions, the spin-orbit orientations of these systems inform the typical outcomes of quiescent planet formation, as well as the primordial stellar obliquity distribution. However, few measurements have been made to constrain the spin-orbit orientations of near-resonant systems. We present a Rossiter-McLaughlin measurement of the near-resonant warm Jupiter TOI-2202 b, obtained using the Carnegie Planet Finder Spectrograph (PFS) on the 6.5m Magellan Clay Telescope. This is the eighth result from the Stellar Obliquities in Long-period Exoplanet Systems (SOLES) survey. We derive a sky-projected 2D spin-orbit angle $\lambda=26^{+12}_{-15}\degree$ and a 3D spin-orbit angle $\psi=31^{+13}_{-11}\degree$, finding that TOI-2202 b -- the most massive near-resonant exoplanet with a 3D spin-orbit constraint to date -- likely deviates from exact alignment with the host star's equator. Incorporating the full census of spin-orbit measurements for near-resonant systems, we demonstrate that the current set of near-resonant systems with period ratios $P_2/P_1\lesssim4$ is generally consistent with a quiescent formation pathway, with some room for low-level ($\lesssim20\degree$) protoplanetary disk misalignments or post-disk-dispersal spin-orbit excitation. Our result constitutes the first population-wide analysis of spin-orbit geometries for near-resonant planetary systems.
\end{abstract}

\vspace{-10mm}

\section{Introduction} 
\label{section:intro}

The dynamical histories of planetary systems can, to some extent, be reconstructed through their current orbital demographics. Near-resonant systems, in which two or more planets exhibit near-exact integer ratio commensurabilities of their orbital periods, offer an especially well-constrained lens into the evolution of planetary systems \citep[e.g.][]{goldreich1965explanation, lee2002dynamics, millholland2018new, goyal2023enhanced}. In these cases, formation models must jointly account for both the systems' near-resonant configurations and their currently observed orbital geometries.  %Here we define near-resonant systems as those with a near-exact integer ratio commensurability in orbital periods, noting that a ``true'' resonance requires the two planets to additionally satisfy the criterion that a critical angle librates about a fixed point.

One observable constraint on a system's orbital geometry is the tilt of companion planets' orbits relative to the host star's spin axis. The sky-projection of these ``spin-orbit'' angles, $\lambda$, can be measured for transiting planets through the Rossiter-McLaughlin effect \citep{rossiter1924detection, mcLaughlin1924some}, in which small radial velocity (RV) shifts are observed across a transiting exoplanet's passage in front of its host star. As the planet transits, it sequentially blocks different red- and blue-shifted components of the stellar disk, leading to a warped signal in the net observed Doppler shift across the transit. The profile of this warped signal encodes the degree of alignment between the planet's transit path and the equator of the host star.

%To date, very few near-resonant systems have had spin-orbit measurements made to characterize the tilts of their constituent planetary orbits. Only two such measurements have been previously obtained. The first was made across a transit of the Kepler-9 b hot Jupiter \citep[$\lambda=-8.2^{+8.7}_{-9.7}\degree$;][]{wang2018stellar}, which lies near a 2:1 resonance with the Kepler-9 c sub-Saturn-mass companion planet \citep{holman2010kepler}. The second was made across a transit of the WASP-148 b hot Jupiter \citep[$\lambda=-13\pm16\degree$;][]{wang2022aligned}, which lies near a 4:1 resonance with the outer WASP-148 c warm Jupiter companion \citep{hebrard2020discovery}. 

%The aligned nature of each of these systems suggests that near-resonant systems may be associated with relatively quiescent formation pathways. These quiescent pathways may have preserved both the near-resonance of the planetary chains and the small spin-orbit angles expected for most protoplanetary disks [cite]. However, with such a small sample size, robust conclusions cannot yet be drawn on a more general scale.

To date, only a handful of near-resonant systems have had spin-orbit angles measured to characterize the tilts of their constituent planetary orbits. In this work, we add a new measurement to this sample: a Rossiter-McLaughlin observation across a transit of the warm Jupiter TOI-2202 b. This is the eighth result from the Stellar Obliquities in Long-period Exoplanet Systems (SOLES) survey \citep{rice2021soles, wang2022aligned, rice2022tendency, rice2023qatar6, hixenbaugh2023spin, dong2023toi, wright2023soles}, which examines the spin-orbit angles of relatively wide-orbiting transiting exoplanets. 

%This is primarily an outcome of the overall bias toward hot Jupiters in the census of spin-orbit measurements, motivated by the deep and frequent transits of hot Jupiters that ease measurement through the Rossiter-McLaughlin effect. However, the broader set of transiting planets discovered by the TESS spacecraft has recently extended the sample of planets amenable to spin-orbit measurements, enabling an extension of spin-orbit constraints to other planet populations -- including the set of resonant and near-resonant systems.

TOI-2202 b lies in a near-resonant configuration deduced through observations of strong transit timing variations (TTVs) from an expected outer companion near the 2:1 mean-motion resonance \citep[MMR;][]{trifonov2021pair}. Using newly obtained RV measurements from the Carnegie Planet Finder Spectrograph \citep[PFS;][]{crane2006carnegie, crane2008carnegie, crane2010carnegie}, together with archival data and new photometry from an assortment of telescopes, we derive a moderate 2D spin-orbit angle $\lambda=26^{+12}_{-15}\degree$ and 3D spin-orbit angle $\psi=31^{+13}_{-11}\degree$ for TOI-2202 b. 

%\citet{goyal2023enhanced} found that planetary systems that include a first-order mean-motion resonances demonstrate an enhanced size uniformity relative to their non-resonant counterparts, favoring a quiescent formation pathway for near-resonant systems. 

Combining our measurement with archival results, we conduct the first population study of the spin-orbit configurations of exoplanets with near-resonant companions. Our findings support the hypothesis that near-resonant planetary systems typically form quiescently -- that is, within the disk plane and without violent post-disk-dispersal interactions, such as planet-planet scattering \citep{rasio1996dynamical}, that would significantly displace orbits from this initial plane -- while simultaneously suggesting the prevalence of low-level dynamical excitation even in near-resonant systems.

%Previous results from this survey include measurements constraining the spin-orbit alignment of the WASP-148 b hot Jupiter with a near-resonant companion \citep{wang2022aligned}; demonstrating the spin-orbit alignment of TOI-1478 b and an emerging trend toward alignment in single-star warm Jupiter systems \citep{rice2022tendency}; showing a joint spin-orbit and orbit-orbit alignment in the Qatar-6 AB b binary star system \citep{rice2023qatar6}; and demonstrating a spin-orbit misalignment for the TOI-1842 b puffy sub-Saturn planet (Hixenbaugh et al., in review).

%Given that TOI-2202 c remains only inferred, but its orbital properties are not yet confirmed, it is not yet possible to distinguish whether the planet pair falls within or simply near a resonance.

\section{Observations}
\label{section:observations}

\subsection{Photometric monitoring}
\begin{figure}
    \centering
    \includegraphics[width=0.48\textwidth]{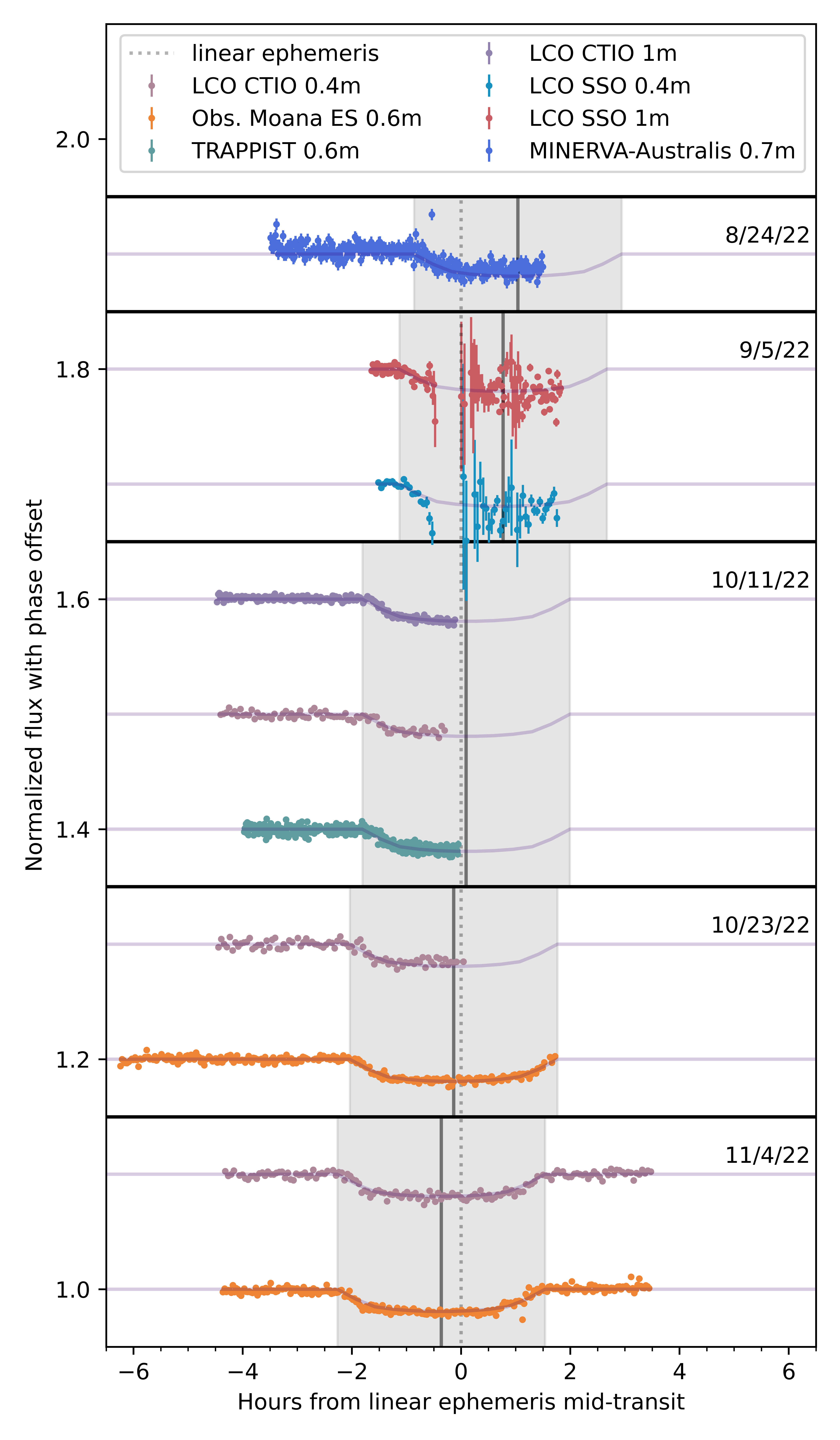}
    \caption{Photometry obtained from the seven telescopes used within this work. Observations are ordered by the date on which they were taken, with shaded regions signifying the modeled transit durations and solid gray lines denoting the transit mid-times measured for each dataset. The transit models associated with our parameter solution are shown together with the data, and a dotted line marks the linear ephemeris mid-transit. The data behind this figure is provided in the digital version of this manuscript.}
    \label{fig:photometry_overview}
\end{figure}

Because TOI-2202 b exhibits strong TTVs, we obtained several photometric transit observations leading up to the Rossiter-McLaughlin event to determine the optimal observing window. We also obtained simultaneous photometry during the scheduled Rossiter-McLaughlin observation to precisely constrain the transit mid-time. Data from seven ground-based telescopes, described in the following subsections and shown in Figure \ref{fig:photometry_overview}, were used in this transit monitoring effort. Each transit of TOI-2202 b lasts 3.8 hours; however, most of our photometric monitoring included observations only at ingress to demonstrate the moving location of the transit start time. The derived transit mid-times associated with each set of observations are provided in Table \ref{tab:ttv_data} for reference.

% \begin{deluxetable}{lllll}
% \tablecaption{Transit numbers $N$, mid-transit times $t_0$, uncertainties $\sigma_{t_0}$, and offsets from the linear ephemeris prediction $\Delta_{\mathrm{linear}}$ for TOI-2202 b based on collected photometry within this work.\label{tab:ttv_data}}
% \tabletypesize{\scriptsize}
% \tablehead{
% \colhead{$N$} & \colhead{$t_0$ (BJD)} & \colhead{$\sigma_{t_0}$} & \colhead{$\Delta_{\mathrm{linear}}$ (min)} & \colhead{Instrument}}
% \tablewidth{300pt}
% \startdata
% 1 & 2459816.26646 & - &  & MINERVA-Australis 0.7m \\
% 2 & 2459828.16992 & - & - & LCO SSO 0.4m \\
% 2 & 2459828.16992 & - & - & LCO SSO 1m \\
% 5 & 2459863.91030 & - & - & LCO CTIO 0.4m \\
% 5 & 2459863.91030 & - & - & LCO CTIO 1m \\
% 5 & 2459863.91030 & - & - & TRAPPIST 0.6m \\
% 6 & 2459875.82376 & - & - & LCO CTIO 0.4m \\
% 6 & 2459875.82376 & - & - & El Sauce 0.6m \\
% 7 & 2459887.73722 & - & - & LCO CTIO 1m \\
% 7 & 2459887.73722 & - & - & El Sauce 0.6m \\
% \enddata
% \end{deluxetable}

\begin{deluxetable}{lllll}
% \tablecaption{Transit numbers $N$, mid-transit times $t_0$, uncertainties $\sigma_{t_0}$, and offsets from the linear ephemeris prediction $\Delta_{\mathrm{lin}}$ for TOI-2202 b based on collected photometry within this work .\label{tab:ttv_data}}
\tablecaption{Transit numbers $N$, mid-transit times $t_0$, uncertainties $\sigma_{t_0}$, and offsets from the linear ephemeris prediction $\Delta_{\mathrm{lin}}$ for TOI-2202 b based on our collected photometry and new TESS transits collected since the publication of \citet{trifonov2021pair}. All listed $\Delta_{\mathrm{lin}}$  values in this table are provided relative to our derived transit mid-time epoch $T_0=2459577.9736362\pm0.0039$ days (Table \ref{tab:results}). \label{tab:ttv_data}}
\tabletypesize{\scriptsize}
\tablehead{
\colhead{$N$} & \colhead{$t_0$ (BJD$_{\rm TDB}$)} & \colhead{$\sigma_{t_0}$} & \colhead{$\Delta_{\mathrm{lin}}$ (min)} & \colhead{Instrument}}
\tablewidth{300pt}
\startdata
-45 & 2459041.9141860 & 0.0013 & 44.91 & TESS \\ 
-44 & 2459053.8287755 & 0.0012 & 46.98 & TESS \\ 
-43 & 2459065.7370344 & 0.0010 & 39.93 & TESS \\ 
-42 & 2459077.6478578 & 0.0010 & 36.58 & TESS \\ 
-41 & 2459089.5520912 & 0.0014 & 23.74 & TESS \\ 
-39 & 2459113.3622859 & 0.0016 & 0.54 & TESS \\ 
-24 & 2459292.0300834 & 0.0013 & -41.90 & TESS \\ 
-23 & 2459303.9535926 & 0.0015 & -26.98 & TESS \\ 
-18 & 2459363.5678775 & 0.0017 & 42.90 & TESS \\ 
-16 & 2459387.4048621 & 0.0016 & 58.28 & TESS \\ 
20 & 2459816.2811297 & 0.0035 & 62.34 & \textsc{Minerva}-Australis \\
21 & 2459828.1830501 & 0.0029 & 46.16 & LCOGT SSO 0.4m \\
& & & & LCOGT SSO 1m \\
24 & 2459863.8942826 & 0.0020 & 5.52 & LCOGT CTIO 0.4m \\
& & & & LCOGT CTIO 1m \\
& & & & TRAPPIST-South  \\
25 & 2459875.7978660 & 0.0014 & -8.25& LCOGT CTIO 0.4m \\
& & & & Observatoire Moana ES \\
26 & 2459887.7015528 & 0.0006 & -21.88 & LCOGT CTIO 1m \\
& & & & Observatoire Moana ES \\
36 & 2460006.8115743 & 0.0012 & -52.83 &TESS\\
37 & 2460018.7353190 & 0.0008 & -37.58 &TESS\\
38 & 2460030.6594249 & 0.0011 & -21.80 &TESS\\
\enddata
\end{deluxetable}

\subsubsection{MINERVA-Australis 0.7m photometry}
Our team measured the ingress of TOI-2202 b's transit on UT 8/24/2022 using one of the 0.7-m \textsc{Minerva}-Australis telescopes \citep{addison2019minerva} located at the University of Southern Queensland's Mount Kent Observatory. The telescope is equipped with a $2000\times2000$ pixel Andor CCD with pixel scale $0.608\arcsec$, and we used a 15-pixel radius ($9.12\arcsec$) aperture to extract the photometry. We obtained $2.65$ hours of pre-transit baseline observations, as well as photometry during the first $2.25$ hours of transit. Observations consisted of continuous 60-second broadband exposures. From this set of observations, we derived a 62.34-minute late ingress of the transit relative to this work's fitted linear ephemeris of TOI-2202 b.

\subsubsection{LCOGT SSO 0.4-m and 1-m photometry}
We measured one partial transit of TOI-2202 b on UT 9/5/2022 using the Las Cumbres Observatory Global Network \citep[LCOGT;][]{brown2013cumbres} Siding Spring Observatory (SSO) 0.4-m and 1-m telescopes, in New South Wales in Australia. Observations were taken in the Sloan $i^{\prime}$ band with 170-second exposures on the 0.4-m telescope and 43-second exposures on the 1-m telescope. Photometry was extracted using the \texttt{AstroImageJ} software \citep{collins2017astroimagej}.

We obtained 0.38 and 0.51 hours of pre-transit baseline observations, as well as photometry during the first 2.88  and 2.95 hours of transit, with the 0.4-m and 1-m LCOGT SSO telescopes, respectively. Conditions were poor during this set of observations, impacting data obtained from both telescopes. Because transparency losses became significant only 30 minutes after the observed ingress, we were able to resolve a clear ingress that occurred $46.16$ minutes late relative to this work's fitted linear ephemeris of TOI-2202 b.

\subsubsection{LCOGT CTIO 0.4-m and 1-m photometry}
\label{subsubsection:lco_ctio}
We also measured two transit ingress events for TOI-2202 b on UT 10/11/2022 and 10/23/2022, as well as one full transit on UT 11/4/2022, using the LCOGT Cerro Tololo Inter-American Observatory (CTIO) 0.4-m telescope located 80 km east of La Serena, Chile. The UT 10/11/2022 ingress was simultaneously observed using the LCOGT CTIO 1-m telescope. Observations were taken in the Sloan $i^{\prime}$ band, with 170-second exposures for the 0.4-m telescope observations and 43-second exposures for the 1-m telescope observations. Photometry was extracted using the \texttt{AstroImageJ} software \citep{collins2017astroimagej}.

From the 0.4-m LCOGT CTIO telescope, we obtained 2.60, 2.41, and 2.05 hours of pre-transit photometry observations, as well as 1.50, 2.08, and 5.74 hours of post-ingress photometry observations during the evenings of UT 10/11/2022, 10/23/2022, and 11/4/2022, respectively. Our observations with the 1-m LCOGT CTIO telescope on UT 10/11/2022  included 2.66 hours of pre-transit photometry observations, as well as 1.69 hours of in-transit data.

Based on the obtained LCOGT CTIO photometry, we found that the 10/11/2022 transit event of TOI-2022 b occurred 5.52 minutes late relative to this work's fitted linear ephemeris of TOI-2202 b. The 10/23/2022 transit event occurred 8.25 minutes early, while the 11/4/2022 event occurred $21.88$ minutes early.

\subsubsection{TRAPPIST-South 0.6-m photometry}
We measured one transit egress event for TOI-2202 b on UT 10/11/2022 using the TRAPPIST-South 0.6-m robotic telescope \citep{jehin2011trappist, Gillon2011} at La Silla Observatory in the Atacama Desert of Chile. Continuous 30-second bservations were taken with the Astrodon ``I+z'' filter. The TRAPPIST-South observing sequence spanned $2.16$ hours prior to the transit ingress, as well as $1.76$ in-transit hours of observations. The observed transit ingress time was consistent with the times derived from the simultaneous LCOGT CTIO 0.4-m and 1-m observations taken on the same night (Section \ref{subsubsection:lco_ctio}), with an ingress $5.52$ minutes after the linear ephemeris prediction.

\subsubsection{Observatoire Moana -- El Sauce 0.6-m photometry}
Our team measured two full transits of TOI-2202 b on UT 10/23/2022 and UT 11/4/2022 using the station of the Observatoire Moana located in El Sauce (ES) Observatory \citep{ropert2021observatorio} in Chile. This station consists of a 0.6m CDK robotic telescope coupled to an Andor iKon-L deep depletion $2000\times2000$ CCD with a scale of $0.67\arcsec$ per pixel. The second of these transits, on UT 11/4/2022, was obtained simultaneously with the presented Rossiter-McLaughlin measurement across the transit of TOI-2202 b. Observations were taken in the Sloan $r$ band with continuous 100-second exposures. 

The 10/23/2022 observations included nearly the full planet transit ($3.76$ in-transit hours), as well as $4.20$ hours of pre-transit baseline photometry. From this dataset, together with the LCOGT CTIO 1-m telescope observation described in Section \ref{subsubsection:lco_ctio}, we derived an 8.25-minute early ingress. The 11/4/2022 observation included $2.10$ hours of pre-transit and $1.91$ hours of post-transit baseline observations, together with continuous observations throughout the transit itself.

\subsection{Radial velocity observations}
We observed the Rossiter-McLaughlin effect across one full transit of TOI-2202 b, from UT 00:48-8:35 on November 4th, 2022, using the Carnegie Planet Finder Spectrograph \citep{crane2006carnegie, crane2008carnegie, crane2010carnegie} on the 6.5m Magellan Clay telescope at Las Campanas Observatory in the southern Atacama Desert of Chile. Our team obtained 25 RV measurements, each with an exposure time of 1100 seconds, $3\times3$ binning, and typical RV precision $\sim3.1$ m/s. Conditions were good throughout the observation, with typical seeing $0.7-0.8\arcsec$ and a small spike in seeing about an hour before transit. The airmass ranged from $z=1.40-1.77$ through the observing sequence. 

In addition to the transit itself, the observing sequence included 1.99 hours of pre-transit and 1.72 hours of post-transit baseline observations. The PFS RV measurements obtained for this work are provided in Table \ref{tab:rv_data} and shown in Figure \ref{fig:rv_joint_fit}.

\begin{deluxetable}{rrrrr}
\tablecaption{PFS RV measurements for the TOI-2202 system, obtained across the transit of TOI-2202 b.\label{tab:rv_data}}
\tabletypesize{\scriptsize}
\tablehead{
\colhead{Time (BJD$_{\rm TDB}$)} & \colhead{RV (m/s)} & \colhead{$\sigma_{\rm RV}$ (m/s)}}
\tablewidth{300pt}
\startdata
2459887.54038&2.96&3.65\\
2459887.55339&-5.13&3.14\\
2459887.56648&0.41&3.30\\
2459887.57971&-9.69&3.28\\
2459887.59266&4.42&3.13\\
2459887.60567&-1.81&2.98\\
2459887.61865&7.04&2.89\\
2459887.63136&0.10&2.94\\
2459887.64460&16.42&2.75\\
2459887.65761&23.90&3.05\\
2459887.67055&8.11&3.12\\
2459887.68319&10.25&3.00\\
2459887.69672&0.13&3.52\\
2459887.70952&-6.93&3.18\\
2459887.72273&-15.10&3.30\\
2459887.73584&-21.73&3.02\\
2459887.74891&-31.60&3.47\\
2459887.76152&-22.51&2.95\\
2459887.77497&-8.69&3.75\\
2459887.78803&3.99&2.96\\
2459887.80100&-3.67&3.26\\
2459887.81431&1.06&2.99\\
2459887.82690&-16.80&3.00\\
2459887.84022&-13.39&3.23\\
2459887.85314&-12.86&3.33\\
\enddata
\end{deluxetable}

% \section{Stellar Parameters} 
%\label{section:stellarparameters}
%[need to get spectrum from Paul for this section? should we even bother with this, since we have no new template data that hasn't been previously published? looks like trifonov+ 2021 used their harps spectrum for stellar parameters, although they had pfs available]

% We used a customized version of \texttt{allesfitter} to fit stellar and planetary parameters simultaneously.  

% \texttt{isochrones } \cite{Morton2015}

% We constructed spectral energy distributions (SEDs) from 2MASS \citep{Cutri2003}, WISE \citep{Cutri2013}, and Gaia \citep{brown2022gaiadr3}. 

% To derive stellar parameters, we conducted model-grid-based inference of stellar parameters for MIST model using \texttt{isochrones} \citep{Morton2015}. We constructed spectral energy distributions (SEDs) using photometry measurements from 2MASS \citep{Cutri2003}, WISE \citep{Cutri2013}, and Gaia \citep{brown2022gaiadr3}. In order to further constrain the stellar radius, we adopted Gaussian priors on Gaia parallax and the upper limit V-band extinction. Their priors are from Gaia DR3 and the 2D dust map provided by \cite{Schlafly2011}. The priors on \teff, system age, and [Fe/H] are drawn from \cite{trifonov2021pair}.

% Gaia parallax from Gaia DR3 \cite{brown2022gaiadr3}

% \section{Obliquity Modeling} 
% \clearpage
\section{Global fitting} 
\label{section:spinorbitmodel}

\subsection{Stellar parameters}
The \texttt{EXOFASTv2} Python package \citep{Eastman2017, Eastman2019} was applied to derive precise stellar parameter values by fitting the host star's spectral energy distribution (SED) using MESA Isochrones \& Stellar Tracks (MIST; \citealt{Choi2016mist,Dotter2016mist}) models. We constructed the SED using photometric measurements from 2MASS \citep{Cutri2003}, WISE \citep{Cutri2013}, and \textit{Gaia} DR3 \citep{brown2022gaiadr3}. Priors on stellar mass (\mstar), age, and effective temperature (\teff) were adopted from \cite{trifonov2021pair}, while priors on parallax ($\varpi$) and upper limits on V-band extinction ($A(V)$) were drawn from \textit{Gaia} DR3 and \cite{Schlafly2011}, respectively. 

As noted by \citet{Eastman2022}, incorporating transit-based densities can provide further constraints on stellar radius and mass, surpassing the limits of systematic error floors on stellar parameters \citep{Tayar2020}. Therefore, we included an additional likelihood term for stellar density in the stellar parameter fit to account for constraints imposed by the transit fit. The output stellar parameters are provided in the top section of Table \ref{tab:results}. All derived stellar parameters are consistent with values reported in \citet{trifonov2021pair} within $1\sigma$.

\subsection{Planetary system parameters}
We applied the \texttt{allesfitter} Python package \citep{gunther2021allesfitter} to refine the TOI-2202 planetary system parameters and derive the sky-projected spin-orbit angle of TOI-2202 b. Using \texttt{allesfitter}, we jointly modeled all photometry and RV observations obtained in this work, together with ten sectors of 30-minute cadence TESS \citep{ricker2015tess} photometry (Sectors 1, 2, 6, 9, 13, 27, 28, 29, 36, 39,  62, and 63)\footnote{TESS data used in this paper can be found in MAST: \dataset[10.17909/0cp4-2j79]{http://dx.doi.org/10.17909/0cp4-2j79}} and all archival PFS, FEROS, and HARPS RVs and ground-based transits reported in \citet{trifonov2021pair}. 

% All parameters listed in Table \ref{tab:results} were allowed to vary in our global fit, with uniform priors set within the provided bounds.
The fitted parameters include the companion's orbital period ($P_b$), the reference mid-time epoch ($T_{0}$), all individual transit mid-times ($t_{0}$), the cosine of the planetary orbital inclination ($\cos{i_b}$), the planet-to-star radius ratio ($R_{b}/R_{\star}$), the sum of radii divided by the orbital semi-major axis ($(R_{\star}+R_{b})/a_b$), the RV semi-amplitude ($K_b$), the parameterized orbital eccentricity and argument of periastron ($\sqrt{e_b}\,\cos{\,\omega_b}$, $\sqrt{e_b}\,\sin{\,\omega_b}$), the sky-projected spin-orbit angle ($\lambda$), the sky-projected stellar rotational velocity ($v\sin i_{\star}$), and twelve limb-darkening coefficients, with two per photometric band ($q_1$ and $q_2$ for each of LCO, El Sauce, \textsc{Minerva}, TRAPPIST-South, and TESS) and two for the in-transit RV dataset ($q_{1; \mathrm{RM}}$ and $q_{2; \mathrm{RM}}$). The systematic offsets between transit and RV datasets obtained by different instruments were accounted for by fitting and subtracting off a quadratic trend between each dataset. During the fit, the jitter term for each RV dataset was added in quadrature. We also fitted for the error scaling factor for each transit, normalized to the original photometric errors to ensure that only the relative weights are important. Parameters for planet c were held fixed at the values derived in \citet{trifonov2021pair}.

 % jitter item for RV data $(\ln{\sigma_\mathrm{jitter}})$, error scaling factor for photometry ($\ln{\sigma}$),

\begin{figure}
    \centering
    \includegraphics[width=0.48\textwidth]{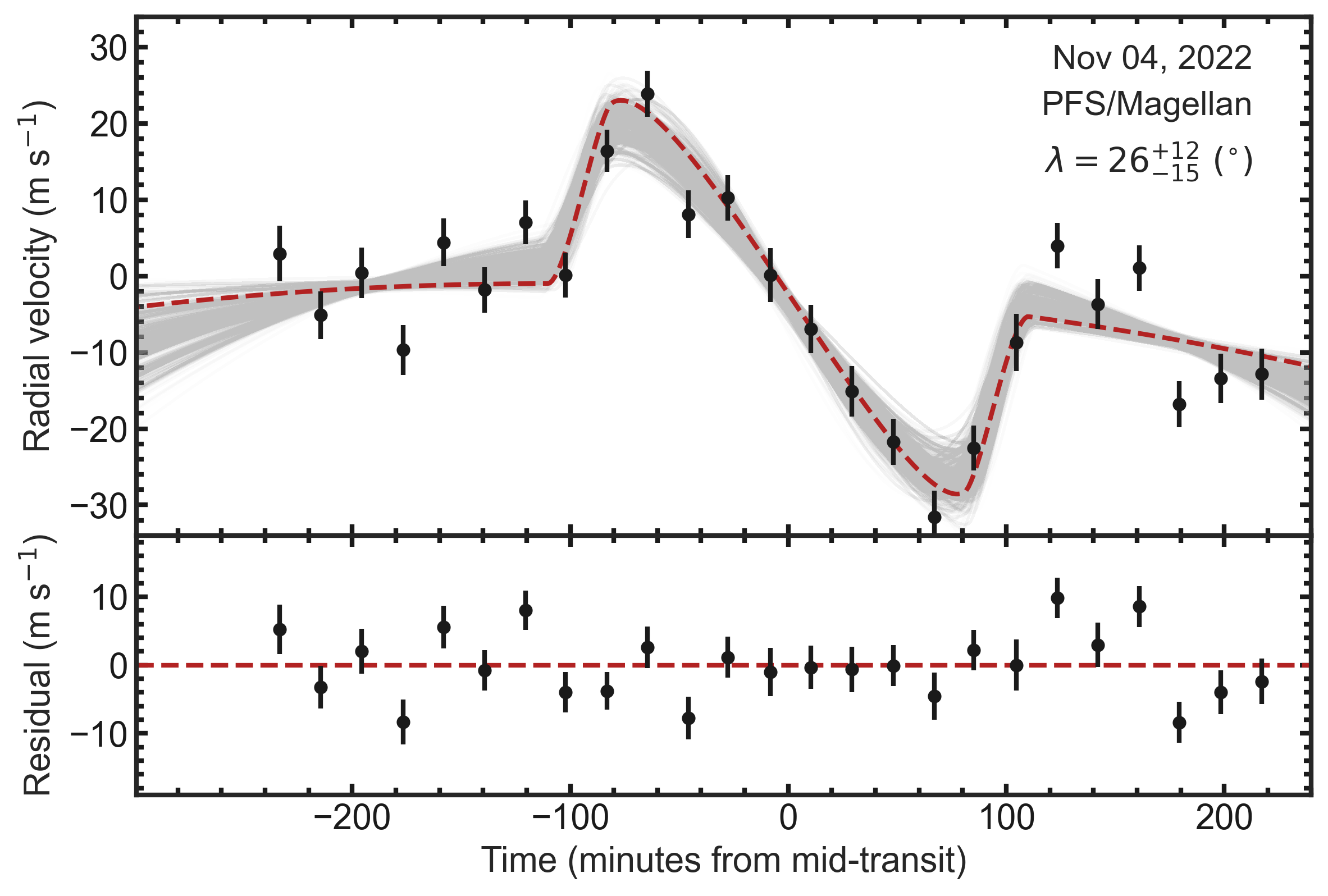}
    \caption{PFS observations of the TOI-2202 b Rossiter-McLaughlin effect from UT 11/4/22, together with the associated uncertainties and best-fitting model (red dashed line). Two thousand Rossiter-McLaughlin model draws from the posterior distribution are shown in gray. Residuals from the best-fitting model are provided in the lower panel.}
    \label{fig:rv_joint_fit}
\end{figure}

Posterior distributions were derived for each free parameter using an affine-invariant Markov Chain Monte Carlo (MCMC) analysis, with 100 walkers that were each run to at least 30$\times$ the autocorrelation length ($\geq500,000$ accepted steps per walker) to ensure convergence. The resulting planetary parameters are provided in Table \ref{tab:results}, while the transit mid-times $t_0$ derived for each light curve are listed in Table~\ref{tab:ttv_data}. The linear ephemeris was derived by applying a weighted least-squares fit to the set of output transit mid-times $t_0$. The reference epoch was optimized to minimize the covariance between $T_{\rm 0}$ and $P_{b}$, and the resulting values are listed in Table~\ref{tab:results}. From this analysis, we derived a moderate sky-projected spin-orbit angle $\lambda=26^{+12}_{-15}\degree$ for TOI-2202 b, with the best-fitting model and residuals shown in Figure \ref{fig:rv_joint_fit}. 

\citet{trifonov2021pair} implemented a Gaussian Process (GP) analysis to derive a stellar rotation period $P_{\mathrm{rot}}=24.1^{+2.3}_{-1.8}$ days for TOI-2202 based on TESS light curve data from Sectors 1, 2, 6, 9, and 13. To update this result, we reexamined the rotational period of TOI-2202 using the full set of currently available TESS data from Sectors 1, 2, 6, 9, 13, 27, 28, 29, 36, 39, 62, and 63, collected over a span of 1716 days from July 25th, 2018 to April 6th, 2023. We applied the GP kernels \texttt{SHOTerm} and \texttt{RotationTerm} that are encapsulated within the \texttt{celerite2} Python package \citep{celerite2}.

Four parallel chains were run using the \texttt{PyMC3} Python package \citep{salvatier2016probabilistic} with an acceptance rate of 0.99, where each chain consisted of 10,000 tuning steps and 10,000 draws. Convergence was deemed to have been achieved when the Gelman-Rubin diagnostic \citep[$\hat R $;][]{Gelman1992} fell below 1.01. The resulting stellar rotation period is \prot$=22\pm1$ days, corresponding to a stellar equatorial velocity $v = (2 \pi R_{*})/P_{\rm rot}=1.86\pm0.19$ km/s.

Lastly, we combined $v$ and $v\sin i_*$ to derive the stellar inclination $i_*$ and the true stellar obliquity $\psi$ for TOI-2202. The Bayesian inference method described in \citet{Masuda2020stincl} and \cite{hjorth2021backward} was applied to account for the interdependent parameters $v$ and $v\sin i_*$, and uniform priors were adopted for the three input parameters $R_*$, $P_{\mathrm{rot}}$, and $\cos i$. This analysis yielded a stellar inclination estimate $i_*=89.77\pm16.76\degree$. Then, the true stellar obliquity ($\psi$) was derived through Equation 9 of \citet{Fabrycky2009},
\begin{equation}
    \cos \psi=\cos i_{*} \cos i+\sin i_{*} \sin i \cos \lambda,
\label{eq:obliquity}
\end{equation}
where $i_{*}$ is the stellar inclination and $i$ is the planet's orbital inclination. The resulting true stellar obliquity is $\psi=31^{+13}_{-11}\degree$.

\begin{deluxetable*}{lllll}
\tablecaption{Priors and posteriors for the TOI-2202 planetary system.\label{tab:results}}
%\tabletypesize{\scriptsize}
\tablehead{}
\tablewidth{300pt}
\startdata
& Description (units) & Priors   & Fitted Value \\
\hline
%& HIRES Spectrum & Alsubai+ 2018 & RM fit \\
%& The Cannon & [fill in] & \texttt{allesfitter} \\ \hline
\\
\multicolumn{5}{l}{Stellar Parameters:}\\
$M_*$  & Mass (\msun)  & - & $0.841^{+0.034}_{-0.032}$\\ \
$R_*$  & Radius (\rsun)  & - & $0.808^{+0.024}_{-0.022}$\\ \
$T_{\rm eff}$  & Effective temperature (K)  & $\mathcal N(5144,104)$ & $5169^{+80}_{-78}$\\ \
$\log{g}$  & Surface gravity (cgs)  & $\mathcal N(4.55,0.20)$ & $4.548^{+0.030}_{-0.031}$\\ \
$[{\rm Fe/H}]$  & Metallicity (dex)  & - & $0.059\pm0.043$\\ \
$v \sin{i_*}$& Projected rotational velocity (km/s)      & $\mathcal U(1.7;0;10)$   &     $2.14_{-0.25}^{+0.28}$      & \\ \
Age  & Age (Gyr)  & - & $6.4^{+4.4}_{-3.9}$\\ \
$A_V$  & V-band extinction (mag)  & $\mathcal U(0.242,0.054)$ & $0.282\pm0.047$\\ \
$\varpi$  & Parallax (mas)  & $\mathcal U(4.226,0.019)$ & $4.225^{+0.016}_{-0.017}$\\ \
$d$  & Distance (pc)  & - & $236.69^{+0.95}_{-0.90}$\\ \
\\
\multicolumn{5}{l}{Planetary Parameters:}\\
$R_b / R_*$&     Planet-to-star radius ratio&  $\mathcal U(0.125;0;1)$   &  $0.1265_{-0.0018}^{+0.0017}$    & \\
$(R_* + R_b) / a_b$&     Sum of radii divided by orbital semimajor axis&   $\mathcal U(0.043;0;1)$   &   $0.04336_{-0.00056}^{+0.00059}$      & \\
$\cos{i_b}$&   Cosine of the orbital inclination&    $\mathcal U(0.011;0;1)$   & $0.0125_{-0.0029}^{+0.0020}$ & \\
$P_b$& Orbital period (days)      & $\mathcal U(11.913;10.913;12.913)$   &    $11.9126075\pm0.00011$ & \\
$T_{0;b}$ & Mid-time epoch (BJD$_{\rm TDB}$) - 2458000 (days) & $\mathcal U(1578;1577;1579)$   &    $1577.9736362\pm0.0038$ &\\
$K_b$& Radial velocity semi-amplitude (km/s)      & $\mathcal U(0;0;10)$   &     $0.0907_{-0.0100}^{+0.0083}$      & \\
$\sqrt{e_b} \cos{\omega_b}$& Eccentricity parameter 1 & $\mathcal U(0;-1;1)$   &     $-0.059_{-0.075}^{+0.089}$      & \\
$\sqrt{e_b} \sin{\omega_b}$& Eccentricity parameter 2 & $\mathcal U(0;-1;1)$   &     $-0.04_{-0.13}^{+0.14}$      & \\
$P_c$& Orbital period (days)      & fixed   &     $24.7557$      & \\
$T_{0;c}$& Mid-time epoch (BJD$_{\rm TDB}$) - 2458000 (days)      & fixed   &    327.103     & \\
$K_c$& Radial velocity semi-amplitude (km/s)      & fixed   &     $0.0196$      & \\
$\sqrt{e_c} \cos{\omega_c}$& Eccentricity parameter 1  & fixed   &     $-0.023$      & \\
$\sqrt{e_c} \sin{\omega_c}$& Eccentricity parameter 2 & fixed   &     $-0.099$      & \\
$\lambda$& Sky-projected spin-orbit angle ($\degree$)      & $\mathcal U(0;-180;180)$   &     $26_{-15}^{+12}$      & \\
$q_{1; \mathrm{LCO}}$& Linear limb-darkening coefficient for LCO&  $\mathcal U(0.5;0;1)$   &   $0.64\pm0.24$    & \\
$q_{2; \mathrm{LCO}}$& Quadratic limb-darkening coefficient for LCO&  $\mathcal U(0.5;0;1)$   &   $0.32_{-0.13}^{+0.21}$    & \\
%-----------------
$q_{1; \mathrm{El\,Sauce}}$& Linear limb-darkening coefficient for El Sauce &  $\mathcal U(0.5;0;1)$   &    $0.60_{-0.36}^{+0.29}$    & \\
$q_{2; \mathrm{El\,Sauce}}$& Quadratic limb-darkening coefficient for El Sauce&  $\mathcal U(0.5;0;1)$   &   $0.28_{-0.20}^{+0.31}$   & \\
%-----------------
$q_{1; \mathrm{\textsc{Minerva}}}$& Linear limb-darkening coefficient for \textsc{Minerva}&  $\mathcal U(0.5;0;1)$   &    $0.43_{-0.18}^{+0.24}$    & \\
$q_{2; \mathrm{MINERVA}}$& Quadratic limb-darkening coefficient for \textsc{Minerva}&  $\mathcal U(0.5;0;1)$   &   $0.37_{-0.24}^{+0.32}$   & \\
%-----------------
$q_{1; \mathrm{TRAPPIST}}$& Linear limb-darkening coefficient for TRAPPIST&  $\mathcal U(0.5;0;1)$   &   $0.43_{-0.13}^{+0.19}$    & \\
$q_{2; \mathrm{TRAPPIST}}$& Quadratic limb-darkening coefficient for TRAPPIST&  $\mathcal U(0.5;0;1)$   &   $0.56_{-0.20}^{+0.23}$    & \\
%-----------------
$q_{1;\rm TESS}$& Linear limb-darkening coefficient for TESS&  $\mathcal U(0.5;0;1)$   &   $0.32_{-0.13}^{+0.20}$    & \\
$q_{2;\rm TESS}$& Quadratic limb-darkening coefficient for TESS &  $\mathcal U(0.5;0;1)$   &   $0.45_{-0.18}^{+0.27}$    & \\
$q_{1;\rm PFS, RM}$& Linear limb-darkening coefficient for the PFS RM &  $\mathcal U(0.5;0;1)$   &   $0.72_{-0.28}^{+0.20}$    & \\
$q_{2;\rm PFS, RM}$& Quadratic limb-darkening coefficient for the PFS RM &  $\mathcal U(0.5;0;1)$   &   $0.56_{-0.32}^{+0.28}$    & \\ 
% $\ln{\sigma_\mathrm{jitter; PFS RM}}$ &Jitter term ($\ln{ \mathrm{km/s} }$)& $\mathcal U(-3;-15;0)$& $-5.58_{-0.31}^{+0.26}$ &  \\
\\
\multicolumn{5}{l}{Derived Parameters:}\\
$M_b$& Planetary mass (M$_\mathrm{Jup}$)&  -   &   $0.904_{-0.10}^{+0.087}$    & \\
$R_b$& Planetary radius (R$_\mathrm{Jup}$)&  -   &   $0.977\pm0.016$    & \\
$a_b / R_*$& Semi-major axis over host radius&  -   &   $25.98\pm0.34$   & \\
$i_b$& Inclination ($\degree$) &  -   &   $89.29_{-0.12}^{+0.17}$    & \\
$e_b$& Eccentricity &  -   &   $0.022_{-0.015}^{+0.022}$    & \\
$\omega_b$& Argument of periastron ($\degree$) &  -   &   $212_{-94}^{+59}$    & \\
$T_\mathrm{tot;b}$& Total transit duration (days)&  -   &   $3.768_{-0.030}^{+0.032}$    & \\
$b$& Impact parameter &  -   &   $0.323_{-0.080}^{+0.055}$   & \\
%-----------------
$u_\mathrm{1; \mathrm{LCO}}$ & Linear limb-darkening coefficient 1 for LCO &  -   &   $0.52\pm0.19$    & \\
$u_\mathrm{2; \mathrm{LCO}}$ & Linear limb-darkening coefficient 2 for LCO &  -   &   $0.28_{-0.33}^{+0.27}$    & \\
%-----------------
$u_\mathrm{1; El\,Sauce}$ & Linear limb-darkening coefficient 1 for El Sauce &  -   &   $0.73_{-0.18}^{+0.15}$    & \\
$u_\mathrm{2; El\,Sauce}$ & Linear limb-darkening coefficient 2 for El Sauce &  -   &   $-0.08_{-0.24}^{+0.30}$   & \\
%-----------------
$u_\mathrm{1; \textsc{Minerva}}$ & Linear limb-darkening coefficient 1 for \textsc{Minerva}&  -   &   $0.38_{-0.27}^{+0.42}$   & \\
$u_\mathrm{2; \textsc{Minerva}}$ & Linear limb-darkening coefficient 2 for \textsc{Minerva}&  -   &   $0.38_{-0.27}^{+0.42}$   & \\
%-----------------
$u_\mathrm{1; TRAPPIST}$ & Linear limb-darkening coefficient 1 for TRAPPIST&  -   &   $0.48\pm0.30$  & \\
$u_\mathrm{2; TRAPPIST}$ & Linear limb-darkening coefficient 2 for TRAPPIST&  -   &   $0.17\pm0.37$   & \\
%-----------------
$u_\mathrm{1; TESS}$ & Linear limb-darkening coefficient 1 for TESS&  -   &   $0.51_{-0.14}^{+0.13}$    & \\
$u_\mathrm{2; TESS}$ & Linear limb-darkening coefficient 2 for TESS&  -   &   $0.06_{-0.25}^{+0.28}$    & \\
%-----------------
$u_\mathrm{1; PFS, RM}$ & Linear limb-darkening coefficient 1 for the PFS RM&  -   &   $0.88_{-0.52}^{+0.46}$    & \\
$u_\mathrm{2; PFS, RM}$ & Linear limb-darkening coefficient 2 for the PFS RM&  -   &   $-0.09_{-0.43}^{+0.52}$    & \\
\enddata
\tablenotetext{}{Note: $P_b$ and $T_{0}$ were derived from the weighted least-square fit to the transit mid-times $t_0$.}
\end{deluxetable*}

\section{Population analysis of near-resonant systems}
\label{section:population}

\subsection{The distribution of spin-orbit angles for near-resonant exoplanets}
\label{subsection:sample_population}

To place this measurement into context, we examined the full set of transiting exoplanet systems with (1) a sky-projected spin-orbit measurement and (2) evidence that the transiting planet lies near a low-order mean-motion resonance with a neighboring companion. We initialized our sample by cross-matching the set of all exoplanets with $\lambda$ measurements in the TEPCat catalogue \citep{southworth2011homogeneous} with the set of exoplanets with one or more confirmed planetary companions around the same host star in the NASA Exoplanet Archive's Planetary Systems table \citep{ps_NEXScI}.\footnote{Both catalogues were accessed on 7/20/2023.} Thirty-two planets were identified that fit these two criteria.\footnote{Because WASP-18 c is a contested planet, the WASP-18 system was removed from our sample.}

We also searched for any planets showing clear sinusoidal TTVs attributable to a resonant or near-resonant planetary companion that has not yet been directly confirmed. We identified all planets with (1) a spin-orbit measurement in the TEPCat catalogue and (2) \texttt{ttv\_flag} = True in the NASA Exoplanet Archive. In addition to TOI-2202 b, nine further candidate near-resonant planets were recovered in this manner. However, after closer examination, we concluded that, other than TOI-2202 b, none of the identified TTV planets without confirmed, nearby companions showed compelling sinusoidal signals (see Appendix \ref{appendix:A} for more details). Therefore, only TOI-2202 was added to the initial set of 32 identified systems with confirmed planetary companions.

%To optimize the fidelity of the sample, we excluded spin-orbit angles that were determined to be misaligned through the observation of a partial, rather than full, transit. These included HIP 41378 d \citep{grouffal2022rossiter} and K2-290 b \citep{hjorth2021backward}. 
Next, we identified systems within our sample with compact configurations near low-order resonances. The sample was restricted to include only systems for which the planet with a spin-orbit measurement has a small period ratio $P_{2}/P_{1}\lesssim4$ relative to at least one of its nearest neighbors. This limit was selected for direct comparison with Figure 4 of \citet{fabrycky2014architecture}. Both inner and outer planetary companions were considered, and the default parameter solution orbital periods from the NASA Exoplanet Archive were adopted for all planets. This cut left 19 planets in 16 systems, with properties described in Table \ref{tab:sample_compact}. 

\begin{figure}
    \centering
    \includegraphics[width=0.48\textwidth]{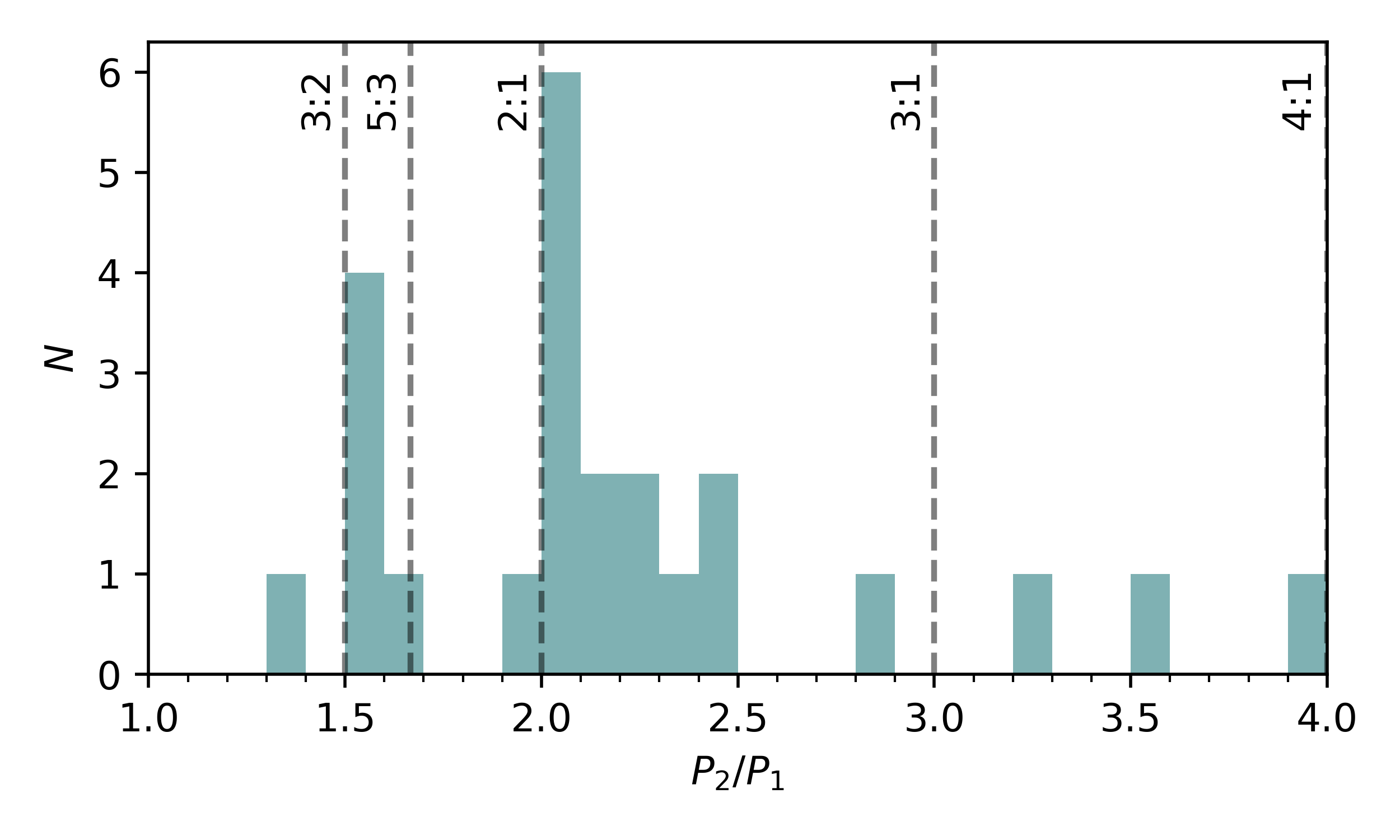}
    \caption{Period ratios of planet pairs in compact systems ($P_{\mathrm{out}}/P_{\mathrm{in}}\lesssim4$) in which at least one planet has a spin-orbit measurement.}
    \label{fig:period_histogram}
\end{figure}

The associated period ratio distribution is shown in Figure \ref{fig:period_histogram}. The inner and outer period ratios for a single planet were included separately in cases where both met the criterion $P_2/P_1\lesssim 4$. For systems in which more than one planet has a spin-orbit measurement (HD 3167, TRAPPIST-1, and V1298 Tau), each planet was separately considered and each relevant period ratio was included only once, for a total of 24 neighboring period ratios with $P_2/P_1\lesssim 4$. 

Lastly, we identified planets within the sample with at least one neighboring companion near a low-order orbital commensurability. Specifically, we searched for planet pairs that fall within 5\% of the 2:1, 3:1, 4:1, 3:2, or 5:3 mean motion commensurabilities. Most near-resonant pairs within the sample were found to lie just wide of the 2:1 and 3:2 resonances (see Figure \ref{fig:period_histogram}), with a distribution comparable to that of the \textit{Kepler} multi-transiting systems examined in \citet{fabrycky2014architecture}. In total, thirteen near-resonances were identified across twelve planet pairs.

\begin{figure*}
    \centering
    \includegraphics[width=0.7\textwidth]{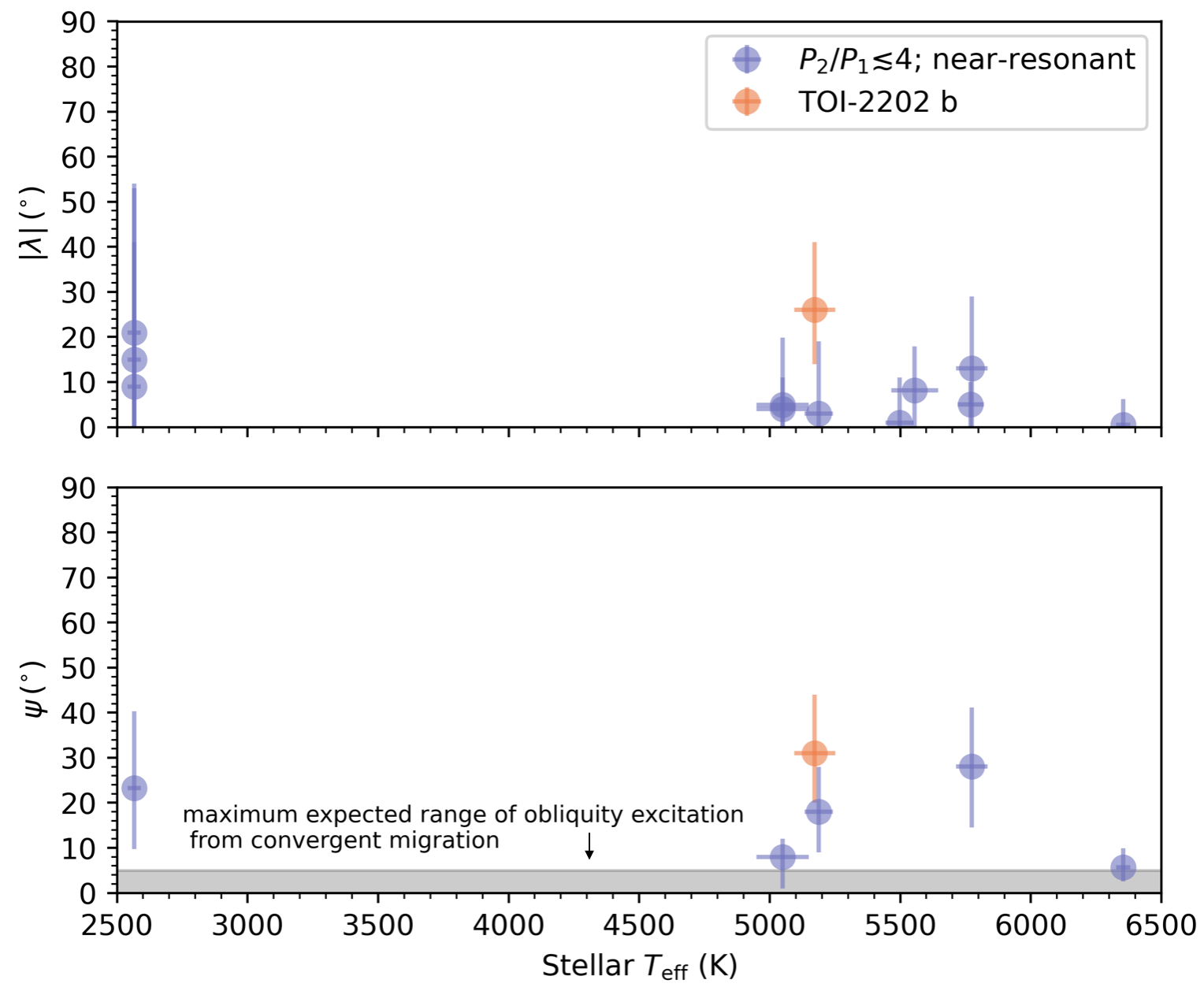}
    \caption{Sky-projected stellar obliquity $|\lambda|$ and 3D stellar obliquity $\psi$ for near-resonant systems. The maximum expected level of stellar obliquity excitation in the convergent migration framework is shown in gray. All $\psi$ values have been measured for separate planetary systems, while V1298 Tau and TRAPPIST-1 include two and three separate displayed $|\lambda|$ values, respectively.}
    \label{fig:teff_v_obliquity}
\end{figure*}

The stellar obliquity distribution for these pairs as a function of host star $T_{\mathrm{eff}}$ is shown in Figure \ref{fig:teff_v_obliquity}. As displayed in the top panel of Figure \ref{fig:teff_v_obliquity} and in Table \ref{tab:sample_compact}, TOI-2202 b is the first exoplanet in a near-resonant configuration for which the measured sky-projected spin-orbit angle has not been consistent with exact alignment ($|\lambda|=0\degree$) within $1\sigma$.

For six near-resonant systems, the 3D spin-orbit angle $\psi$ has also been derived. These systems are shown in the bottom panel of Figure \ref{fig:teff_v_obliquity}. The distribution of 3D angles reveals that a few more systems, in addition to TOI-2202 b, are likely offset from exact alignment.  While no systems with near-resonant configurations have been found with strong misalignments indicative of polar or retrograde orbits, the scatter in the 3D distribution suggests some range in true stellar obliquities even in near-resonant systems.

To quantify this deviation from alignment, we drew 10,000 iterations of random values from each of the six systems with measured $\psi$ values, using the reported Gaussian uncertainties from each measurement. Each iteration was then compared with an ``aligned'' Rayleigh distribution of 100,000 values, with scale parameter $\sigma=1.8$ such that $\sim98$\% of draws fall within the range $\psi<5\degree$. A Kolmogorov-Smirnov (K-S) test quantifying the difference between these two distributions returns $p<0.05$ for 99.4\% of random draws, whereas 0.6\% of random draws return $p>0.05$. As a result, we reject the null hypothesis that the observed $\psi$ distribution exhibits consistent, near-exact ($\psi<5\degree$) alignment.

\subsection{Additional relevant systems}
\label{subsection:additional_relevant_systems}

A few systems were excluded from this population that may also serve to inform the distribution of spin-orbit angles for near-resonant systems. While these systems do not fit the criteria used to develop the uniform sample in Section \ref{subsection:sample_population}, they each offer further relevant insights into the dynamical evolution of near-resonant exoplanet pairs.

One notable omission is a 1D spin-orbit measurement -- obtained in the inclination direction -- that indicates a misalignment in the compact multi-planet system Kepler-56 \citep{huber2013stellar}. This system was excluded because it does not have a reported $\lambda$ constraint. Instead, an asteroseismic analysis conducted by \citet{huber2013stellar} revealed a stellar spin axis inclined at $i_*\sim45\degree$. The Kepler-56 system includes two transiting planets that, by definition, have $i_b\sim i_c\sim90\degree$, such that the stellar spin axis at $i_*\sim45\degree$ indicates a substantial misalignment in the line-of-sight direction. Kepler-56 b and c lie near a 2:1 (c:b) mean motion commensurability.

Another relevant system is 55 Cancri, which is an aligned multiplanet system that includes a near-resonance \citep{mcarthur2004detection, fischer2008five, nelson2014cancri}. This system was not included within our analysis because the only planet with a spin-orbit measurement in the system, 55 Cancri e, has a large period ratio $P_b/P_e\sim 20$ with its nearest neighbor. However, a 3:1 near-resonance exists elsewhere in the system between 55 Cancri b and c, suggesting that this system may have formed in a similar manner to the the systems in our sample. 55 Cancri e has a measured spin-orbit angle $\lambda=10^{+17}_{-20}\degree$ and $\psi=23^{+14}_{-12}\degree$ \citep{zhao2023measured}.

A third relevant system is HIP 41378, with a previously reported misalignment $|\lambda|=57^{+26}_{-18}\degree$ for the planet HIP 41378 d \citep{grouffal2022rossiter}. This system was excluded from the sample because the orbital period of HIP 41378 d has not been precisely confirmed. Only a partial Rossiter-McLaughlin observation has been obtained for this system due to the planet's long transit duration, with measurement uncertainties in the acquired dataset that are comparable to the signal amplitude \citep{grouffal2022rossiter}. HIP 41378 d provides an especially interesting case study as one of the few long-period transiting exoplanets that is amenable to spin-orbit measurements. Additional measurements would be helpful to more clearly establish whether this planet lies within a near-resonance and to more precisely constrain its spin-orbit configuration.

%HIP 41378 (f:d) shows evidence of lying near a 2:1 MMR \citep{grouffal2022rossiter}.\citet{weiss2013mass} showed that Kepler-89 d:c show evidence of lying near a 2:1 MMR.\citet{david2019four} demonstrated that V1298 Tau d:c lie near a 3:2 MMR, while \citet{feinstein2022v1298} found evidence suggesting that V1298 Tau e:b may lie near a 2:1 MMR.

%Two systems that have been found to be misaligned in previous works are included within this sample: HIP 41378 d and HD 3167 c. However, HIP 41378 d was deemed to be misaligned based on a partial transit observation, which may introduce biases into the result. HD 3167 c has not shown a resolved signal with uncertainties smaller than the signal amplitude \citep{dalal2019, bourrier2021}, and the system has a low $v\sin i_*$ measurement, suggesting that models may suffer from an overestimate of $v\sin i_*$ that can be interpreted as a polar planet. Further high-precision observations are necessary to confirm the properties of these two systems.

\begin{deluxetable*}{lcccccccccc}
\tablecaption{Systems with a spin-orbit measurement and $P_2/P_1 \lesssim 4$. $N_{\rm pl}$ is the number of confirmed planets in the system to date. $N_{\rm pl}$, orbital period, and $T_{\rm eff, *}$ have values drawn from the default parameter solution listed in the NASA Exoplanet Archive on 7/20/23. Identified MMRs are provided for commensurabilities in which the listed planet is the outer (longer-period) planet of the pair, as well as those in which the listed planet is the inner (shorter-period) planet of the pair. References are provided for $\lambda$ measurements. $\psi$ values were drawn from \citet{albrecht2021preponderance} when possible for systems with no measurement reported in TEPCat. \label{tab:sample_compact}}
\tabletypesize{\scriptsize}

\tablehead{
\colhead{} & \colhead{System} & \colhead{Planet} & \colhead{$N_{\rm pl}$} & \colhead{Orbital period (days)} & \colhead{MMR$_{\rm out}$} & \colhead{MMR$_{\rm in}$} & \colhead{$T_{\mathrm{eff}, *}$ (K)} & \colhead{$|\lambda|\, (\degr)$} & \colhead{Reference, $|\lambda|$} & \colhead{$\psi\, (\degr)$}}
\tablewidth{300pt}
\startdata
%& 55 Cnc$^{*}$ & e & 5 & $0.7365474^{+0.0000013}_{-0.0000014}$ & - & - & $5172\pm18$ & $10_{-20}^{+17}$ & $23^{+14}_{-12}$ \\
& AU Mic & b & 2 & $8.463000\pm0.000002$ & - & - & $3588\pm87$ & $3.0_{-10.4}^{+10.3}$ & \citet{palle2020transmission} & $12.1^{+11.3}_{-7.5}$\\
%& HD 3167 & b & 4 & $0.959641\pm0.000011$ & - & - & $5261\pm60$ & $6.6_{-6.6}^{+7.9}$ & $29.5^{+7.2}_{-9.4}$ \\
& HD 3167 & c & 4 & $29.8454\pm0.0012$ & - & - & $5261\pm60$ & $108.9_{-5.4}^{+5.5}$ & \citet{bourrier2021rossiter} & $107.7^{+5.1}_{-4.9}$ \\
& HD 63433 & b & 2 & $7.10793^{+0.00040}_{-0.00034}$ & - & - & $5640\pm74$ & $8_{-45}^{+33}$ & \citet{mann2020tess} & $25.6^{+22.5}_{-15.3}$ \\
& HD 106315 & c & 2 & $21.05704\pm0.00046$ & - & - & $6327\pm48$ & $2.7_{-2.7}^{+2.6}$ & \citet{bourrier2023dream} & - \\
%& HIP 41378$^{*}$ & d & 5 & $278.3618$ & - & 2:1 (f:d) & $6226\pm43$ & $57_{-18}^{+26}$ & - \\
& Kepler-9$^{*}$ & b & 3 & $19.23891\pm0.00006$ & - & 2:1 (c:b) & $5774\pm60$ & $13\pm16$ & \citet{wang2018stellar} & $28.1^{+13.0}_{-13.6}$ \\
& Kepler-25$^{*}$ & c & 3 & $12.72070^{+0.00011}_{-0.00010}$ & 2:1 (c:b) & - & $6354\pm27$ & $0.5\pm5.7$ & \citet{albrecht2013low} & $5.7^{+4.2}_{-3.2}$ \\
& Kepler-30$^{*}$ & b & 3 & $29.33434\pm0.00815$ & - & 2:1 (c:b) & $5498\pm54$ & $1\pm10$ & \citet{sanchis2012alignment}& - \\
& Kepler-89 & d & 4 & $22.3429890\pm0.0000067$ & 2:1 (d:c) & - & $6182\pm58$ & $6_{-13}^{+11}$ & \citet{hirano2012planet} & - \\
& TOI-942 & b & 2 & $4.324210\pm0.000019$ & - & - & $4969\pm100$ & $1_{-33}^{+41}$ & \citet{wirth2021toi} & $2^{+27}_{-33}$ \\
& TOI-1136$^{*}$ & d & 6 & $12.51937^{+0.00037}_{-0.00041}$ & 2:1 (d:c) & 3:2 (e:d) & $5770\pm50$ & $5\pm5$ & \citet{dai2023toi} & $<28$ \\
& TOI-2076$^{*}$ & b & 3 & $10.35509^{+0.00020}_{-0.00014}$ & - & 2:1 (c:b) & $5187_{-53}^{+54}$ & $3_{-15}^{+16}$ & \citet{frazier2023neid} & $18^{+10}_{-9}$ \\
& TOI-2202$^{*}$ & b & 2 & $11.9126075\pm0.00011$ & - & 2:1 (c:b) & $5169^{+80}_{-78}$ & $26^{+15}_{-12}$ & this work & $31^{+13}_{-11}$\\
& TRAPPIST-1$^{*}$ & b & 7 & $1.510826\pm0.000006$ & - & 4:1 (e:b) & $2566\pm26$ & $15_{-30}^{+26}$ & \citet{hirano2020evidence}  & $^{\dag}23.3^{+17.0}_{-13.6}$ \\
& TRAPPIST-1$^{*}$ & e & 7 &$6.101013\pm0.000035$ & 3:2 (e:d) & 3:2 (f:e) & $2566\pm26$ & $9_{-51}^{+45}$ & \citet{hirano2020evidence} & $^{\dag}23.3^{+17.0}_{-13.6}$ \\
& TRAPPIST-1$^{*}$ & f & 7 & $9.207540\pm0.000032$ & 3:2 (f:e) & 2:1 (h:f) & $2566\pm26$ & $21\pm32$ & \citet{hirano2020evidence} & $^{\dag}23.3^{+17.0}_{-13.6}$ \\
& V1298 Tau$^{*}$ & b & 4 & $24.1396\pm0.0018$ & 2:1 (b:d) & 2:1 (e:b) & $5050\pm100$ & $4_{-10}^{+7}$ & \citet{johnson2022aligned} & $8^{+4}_{-7}$ \\
& V1298 Tau$^{*}$ & c & 4 & $8.24958\pm0.00072$ & - & 3:2 (d:c) & $5050\pm100$ & $4.9_{-15.1}^{+15.0}$ & \citet{feinstein2021h} & - \\
& WASP-47 & b & 4 & $4.1591492\pm0.0000006$ & - & - & $5552\pm75$ & $0\pm24$ & \citet{sanchis2015low} & $29^{+11}_{-13}$ \\
& WASP-148$^{*}$ & b & 2 & $8.803809\pm0.000043$ & - & 4:1 (c:b) & $5555\pm90$ & $8.2_{-8.7}^{+9.7}$ & \citet{wang2022aligned} & - \\
\enddata
\tablenotetext{$*$}{At least one low-order period commensurability has been identified in this system. The TRAPPIST-1 and TOI-1136 systems are each resonant chains, with additional commensurabilities within each system that are not listed in this table.}
\tablenotetext{$\textdagger$}{A single $\psi$ value was derived for the TRAPPIST-1 system in \citet{albrecht2021preponderance}, such that these three values all correspond to just one measurement.}
%\tablenotetext{$\dag$ All}
\end{deluxetable*}

%55 Cnc, AU Mic, HD 3167, HD 63433, HD 106315, HIP 41378, Kepler-9, Kepler-25, Kepler-30, KOI-94, TOI-942, TOI-1136, TOI-2076, TRAPPIST-1, V1298 Tau, WASP-47, WASP-148

%The identified systems included 55 Cnc e, AU Mic b, HD 3167 b, HD 3167 c, HD 63433 b, HD 106315 c, HIP 41378 d, Kepler-9 b, Kepler-25 c, Kepler-30 b, KOI-94 d, TOI-942 b, TOI-1136 d, TOI-2076 b, TRAPPIST-1 b, TRAPPIST-1 e, TRAPPIST-1 f, V1298 Tau b, V1298 Tau c, WASP-47 b, and WASP-148 b. We note that WASP-18 c is a contested planet, and therefore we removed WASP-18 b from our sample. 

\section{Discussion}
\label{section:discussion} 

The measured spin-orbit angle of TOI-2202 b, together with the full census of spin-orbit measurements for near-resonant exoplanets, indicates that even quiescently formed systems may experience low-level dynamical excitation that produces some dispersion in their spin-orbit orientations. The root of this excitation is intertwined with the underlying formation and prevalence of resonances in planetary systems.

At the population level, most near-resonant planets identified by \textit{Kepler} have been observed to lie just wide of true mean motion resonances \citep{lissauer2011architecture, fabrycky2014architecture}. This finding has been generally interpreted as evidence that such systems began in resonant configurations and were later displaced from deep resonance. Within this framework, near-resonant systems constitute a population that has successfully retained the imprints of past dynamical capture into mean motion resonance -- a delicate configuration that easily diverges from exact commensurability through post-disk-dispersal dynamical perturbations \citep[e.g.][]{michtchenko2008dynamic, michtchenko2008dynamic2, deck2012rapid, izidoro2017breaking, leleu2021six} -- such that they offer key clues into their host systems' primordial architectures.

Convergent migration \citep[e.g.][]{goldreich1965explanation, snellgrove2001disc, lee2002dynamics, rein2012period, bitsch2019formation} could feasibly produce mean motion resonances across a broad range of planetary systems. Planet pairs may be gently pushed away from nominal mean motion resonances in the post-disk-dispersal stage through eccentricity damping and orbital circularization \citep{terquem2007migration, lithwick2012resonant, batygin2013dissipative, goldreich2014overstable, delisle2014resonance, chatterjee2015planetesimal} or through in situ mass growth by planets on circular orbits \citep{petrovich2013planets}. %Resonances may, alternatively, be disrupted \textbf{more violently} through dynamical scattering events with a planetesimal disk that drives planet migration \citep{thommes2008mean, chatterjee2015planetesimal}.

%The absence of very large spin-orbit misalignments ($\psi\gtrsim45\degree$) within the observed sample may suggest that systems that maintain configurations near mean-motion resonance have not encountered stellar obliquity excitation through secular resonances or other similar mechanisms. 

%In the case that most near-resonant systems began in true resonant configurations, the same mechanism that serves to break resonances may also produce low-level misalignments. However, within the convergent migration framework \textbf{as modeled by \citet{izidoro2017breaking}}, the spin-orbit distribution of planets formed within aligned protoplanetary disks is expected to peak at $\psi\leq5\degree$ \citep{esteves2023assessing} from the process of resonance disruption alone. Planet migration, likewise, is expected to occur within the plane of the host protoplanetary disk. 

Convergent migration and gentle resonance divergence mechanisms are expected to operate within the plane of the host protoplanetary disk. Even in the more dynamically violent instability framework modeled by \citet{izidoro2017breaking}, the spin-orbit distribution of planets formed within aligned protoplanetary disks is expected to peak at $\psi\leq5\degree$ \citep{esteves2023assessing} from the process of resonance disruption alone. Therefore, nonzero spin-orbit misalignments may trace small primordial tilts of the systems' natal protoplanetary disks. Previous studies examining the mutual inclinations between stellar rotation axes and their surrounding protoplanetary \citep{davies2019star} and debris \citep{hurt2023evidence} disks have each demonstrated evidence consistent with a prevalence of low-level ($\lesssim20\degree$) disk misalignments that could be attributed solely to chaotic accretion in turbulent molecular cloud cores \citep{takaishi2020star}.

Alternatively, low-level misalignments in near-resonant systems may be produced by a mechanism that does not require a primordially misaligned inner disk. For example, \citet{gratia2017outer} showed that planet-planet scattering in the outer region of a planetary system may gently produce small misalignments of up to $\sim20\degree$ in the inner planetary system, offering a possible avenue to tilt a system without disrupting near-resonances. A misaligned outer planet, produced through either planet-planet scattering in the outer system or through a misaligned outer disk \citep[e.g.][]{nealon2019scattered}, could also potentially tilt its inner companions \citep{zhang2021long} while preserving a near-resonant configuration.

The planets in compact, near-resonant systems considered within this work span a range of masses, from sub-Earth-mass \citep[TRAPPIST-1 e at $0.692\pm0.022 M_{\oplus}$;][]{agol2021refining} to Jovian-mass planets (TOI-2202 b at $0.904^{+0.087}_{-0.10} M_J$; this work). This wide range of planet masses may encompass multiple regimes of planet formation and migration that have not been disentangled within our analysis. The current set of observations does not demonstrate a clear distinction between the spin-orbit angles of lower- and higher-mass planets in near-resonant configurations. An expanded sample may unveil population differences, if present, across mass regimes.

%The scarcity of higher-mass, Jupiter analogue planets within this sample, despite their overrepresentation across the census of planets with spin-orbit measurements, may suggest that higher-mass transiting planets do not commonly retain near-resonances -- either due to never entering these configurations, or due to post-formation dynamical disruption within these systems.

Further monitoring of the TOI-2202 system is needed to more precisely pinpoint the properties of the TOI-2202 c planet and to constrain the presence of additional companions within the system. A direct confirmation of TOI-2202 c would offer the opportunity to demonstrate whether the system lies within, or only near, a ``true'' resonance, such that a critical angle in the system librates about a fixed point. More broadly, additional high-precision Rossiter-McLaughlin measurements for near-resonant systems offer a promising path forward to constrain the origins of low-level misalignments in quiescently formed systems.

Across the broader population of exoplanets with spin-orbit measurements -- including those that are \textit{not} in near-resonant systems -- previous work has found evidence for a nonzero mean stellar obliquity with significant scatter $\psi=19\pm10\degree$ \citep{munoz2018statistical}. The apparent persistence of this deviation from alignment, even in near-resonant systems, suggests the universality of low-level dynamical excitation -- a pattern well exemplified by the TOI-2202 system.
%\section{Conclusions}

%We have presented the Rossiter-McLaughlin measurement of the near-resonant TOI-2202 b warm Jupiter together with several epochs of transit photometry leading up to the RM event. Based on this measurement, we have demonstrated that TOI-2202 b lies on a near-aligned orbit relative to its host star, with $\lambda=34^{+16}_{-15}\degree$ and $\psi=xxxx$. Spin-orbit measurements in systems such as TOI-2202 provide important constraints on the range of inclinations spanned by quiescently formed planetary systems.

\label{section:conclusions}

\section{Acknowledgements}
\label{section:acknowledgements}

We thank the anonymous referee for their valuable feedback on this manuscript. We thank Chelsea Huang for her support in organizing observations for this work. M.R. and S.W. thank the Heising-Simons Foundation for their generous support. M.R. acknowledges support from Heising-Simons Foundation Grant \#2023-4478, as well as the 51 Pegasi b Fellowship Program. This paper includes data gathered with the 6.5 meter Magellan Telescopes located at Las Campanas Observatory, Chile. This research has made use of the NASA Exoplanet Archive, which is operated by the California Institute of Technology, under contract with the National Aeronautics and Space Administration under the Exoplanet Exploration Program. A.J. acknowledges support from ANID -- Millennium  Science  Initiative -- ICN12\_009 and FONDECYT project 1210718.  R.B. acknowledges support from FONDECYT project 11200751 and additional support from ANID -- Millennium  Science  Initiative -- ICN12\_009.

This work makes use of observations from the LCOGT network. Part of the LCOGT telescope time was granted by NOIRLab through the Mid-Scale Innovations Program (MSIP). MSIP is funded by NSF. This research has made use of the Exoplanet Follow-up Observation Program (ExoFOP; DOI: 10.26134/ExoFOP5) website, which is operated by the California Institute of Technology, under contract with the National Aeronautics and Space Administration under the Exoplanet Exploration Program. K.A.C. acknowledges support from the TESS mission via subaward s3449 from MIT. 

This work makes use of data from the \textsc{Minerva}-Australis facility. \textsc{Minerva}-Australis is supported by Australian Research Council LIEF Grant LE160100001, Discovery Grants DP180100972 and DP220100365, Mount Cuba Astronomical Foundation, and institutional partners University of Southern Queensland, UNSW Sydney, MIT, Nanjing University, George Mason University, University of Louisville, University of California Riverside, University of Florida, and The University of Texas at Austin. We respectfully acknowledge the traditional custodians of all lands throughout Australia, and recognise their continued cultural and spiritual connection to the land, waterways, cosmos, and community. We pay our deepest respects to all Elders, ancestors and descendants of the Giabal, Jarowair, and Kambuwal nations, upon whose lands the \textsc{Minerva}-Australis facility at Mt Kent is situated.

TRAPPIST-South is funded by the Belgian National Fund for Scientific Research (F.R.S.-FNRS) under grant PDR T.0120.21, with the participation of the Swiss National Science Fundation (SNF). M.G. and E.J. are FNRS Senior Research Associates. The postdoctoral fellowship of K.B. is funded by F.R.S.-FNRS grant T.0109.20 and by the Francqui Foundation. F.J.P. acknowledges financial support from the grant CEX2021-001131-S funded by MCIN/AEI/ 10.13039/501100011033. This publication benefits from the support of the French Community of Belgium in the context of the FRIA Doctoral Grant awarded to MT. This research was supported in part by Lilly Endowment, Inc., through its support for the Indiana University Pervasive Technology Institute.

\software{\texttt{allesfitter} \citep{gunther2021allesfitter}, \texttt{AstroImageJ} \citep{collins2017astroimagej}, \texttt{emcee} \citep{foremanmackey2013}, \texttt{lightkurve} \citep{Lightkurve2018}, \texttt{matplotlib} \citep{hunter2007matplotlib}, \texttt{numpy} \citep{oliphant2006guide, walt2011numpy, harris2020array}, \texttt{pandas} \citep{mckinney2010data}, \texttt{ptemcee} \cite{Vousden2016}, \texttt{scipy} \citep{virtanen2020scipy}}

\facility{Magellan II: PFS; \textsc{Minerva}-Australis 0.7m; LCOGT SSO 0.4m and 1m; LCOGT CTIO 0.4m and 1m; TRAPPIST-South 0.6m; Observatoire Moana -- El Sauce 0.6m; NASA Exoplanet Archive; TEPCat}

\appendix 
\section{Vetting TTV Planets}
\label{appendix:A}

Here, we discuss the vetting process applied in Section \ref{section:population} to identify planets with observed TTVs that are likely indicative of near-resonant companions. The candidate planets with spin-orbit measurements and with TTV signals, but with no confirmed planet attributable to the TTV signal, include CoRoT-2 b, HAT-P-13 b, KELT-19 A b, KOI-12 b, KOI-13 b, Kepler-17 b, Qatar-1 b, WASP-12 b, and WASP-43 b.

None of the identified TTV planets without confirmed, nearby companions included clear sinusoidal signals. WASP-12 b and Qatar-1 b were removed from the sample, since both systems have been monitored and found to show no sinusoidal TTVs \citep{collins2017transit}. We note that the Qatar-1 b TTV detection has also been contested \citep{von2013qatar, maciejewski2015no}. Likewise, WASP-12 b demonstrates transit timing variations that previous studies have found are most consistent with a decaying orbit \citep{yee2020orbit, turner2021decaying, wong2022tess}. The TTV observations of WASP-43 b have also been attributed to orbital decay \citep{jiang2016possible}, which was later ruled out \citep{hoyer2016ruling, garai2021orbit}; however, the WASP-43 b TTVs show no clear signs of periodicity. No clear periodicities were identified in the CoRoT-2 b, HAT-P-13 b, or KELT-19 A b TTVs based on the results reported in \citet{ivshina2022tess}. \citet{holczer2016transit} found that the frequency of Kepler-17 b's TTVs may be attributable to the star's rotational frequency, and that the TTVs of KOI-13 b show a strong stroboscopic effect such that they may not be associated with a companion planet. \citet{holczer2016transit} also identified no clearly sinusoidal periodicity in the TTVs observed for KOI-12 b.

\bibliography{bibliography}
\bibliographystyle{aasjournal}

\end{document}